\documentclass{article}

\usepackage[final,nonatbib]{neurips_2024}

\usepackage[utf8]{inputenc} % allow utf-8 input
\usepackage[T1]{fontenc}    % use 8-bit T1 fonts
\usepackage{hyperref}       % hyperlinks
\usepackage{url}            % simple URL typesetting
\usepackage{booktabs}       % professional-quality tables
\usepackage{amsfonts}       % blackboard math symbols
\usepackage{nicefrac}       % compact symbols for 1/2, etc.
\usepackage{microtype}      % microtypography
\usepackage{xcolor}         % colors
\usepackage{graphicx}
\usepackage[square,numbers]{natbib}
\usepackage{hyperref}
\usepackage{comment}

\title{Asynchronous Hebbian/anti-Hebbian networks}

\author{%
  Henrique Reis Aguiar \\
  Institute for Adaptive and Neural Computation \\
  University of Edinburgh\\
  \texttt{s1430659@ed.ac.uk} \\
   \AND
  Matthias H. Hennig \\
  Institute for Adaptive and Neural Computation \\
  University of Edinburgh\\
  \texttt{m.hennig@ed.ac.uk} \\
}

\begin{document}

\maketitle

\begin{abstract}
     Lateral inhibition models coupled with Hebbian plasticity have been shown to learn factorised causal representations of input stimuli, for instance, oriented edges are learned from natural images. Currently, these models require the recurrent dynamics to settle into a stable state before weight changes can be applied, which is not only biologically implausible, but also impractical for real-time learning systems. Here, we propose a new Hebbian learning ruleewhich is mented withusing plausible biological mechanisms that have been observed experimentally. We find that this rule allows for efficient, time-continuous learning of factorised representations, very similar to the classic noncontinuous Hebbian/anti-Hebbian learning. Furthermore, we show that this rule naturally prevents catastrophic forgetting when stimuli from different distributions are shown sequentially.
\end{abstract}

Published in the First Workshop on NeuroAI@NeurIPS2024.

\section{Introduction}

Neural network models with inhibition and local Hebbian plasticity have been extensively analyzed and shown to learn factorised representations of input data, which manifest in appropriate feedforward weights or receptive fields \citep{foldiak1990forming, savin2010independent, pehlevan2014hebbian, falconbridge2006simple, krotov2019unsupervised}. Factorised receptive fields constitute an efficient representation of sensory data that is hypothesized to be used in sensory systems \cite{barlow1961possible, qin2023coordinated}. It is useful from a computational perspective to represent high-dimensional sensory stimuli in terms of constituent parts or {\em factors} (e.g., edges for natural images) \cite{bell1997independent} as these generalize well across images and space and also support subsequent tasks such as object recognition \cite{bengio2013representation}. A factorised representation is also sparse \cite{olshausen2004sparse} as only a few neurons are active at a certain time, which can reduce the metabolic cost \cite{baddeley1996efficient} and maximize the capacity of a subsequent associative memory \cite{willshaw1969non, baum1988internal}. \

Most lateral inhibition models that learn factorised representations have an important caveat: recurrent dynamics need to reach a stable state before a plasticity update can be applied \cite{pehlevan2014hebbian,krotov2019unsupervised}. This expectation maximization procedure is implausible from a biological perspective as neural dynamics and plasticity evolve continuously, and do not necessarily evolve on very different time scales \cite{linsker2005improved, minden2018biologically}. Particularly in a continuous world one cannot separate (or discretize) a sequence of incoming stimuli, and therefore it is unclear when the weights should actually be updated. \

Networks that do not require recurrent dynamics have previously been proposed \cite{linsker2005improved, minden2018biologically}, however, either they do not learn factorised representations \cite{minden2018biologically} or they cannot maintain weight stability \cite{linsker2005improved}. Other lines of work have applied ongoing plasticity with small learning rates and successfully obtained factorised representations \cite{eckmann2024synapse}, however they still require holding the current stimuli long enough for the recurrent dynamics to settle. \
%Waiting for the recurrent dynamics to reach a fixed point is unrealistic in biological networks since stimuli vary at arbitrary speeds. \

Here we propose a Hebbian plasticity model in which post-synaptic neurons update their incoming weights asynchronously. Each neuron performs an update when its activity reaches a certain threshold. Importantly, we also introduce a refractory period which prevents multiple continuous updates of the same post-synaptic neuron. Such refractoriness in LTP has been observed experimentally \cite{kramar2012synaptic, flores2024synapse}, but to our knowledge has not been incorporated into plasticity models. We show that our model yields factorised representations which models with on-going Hebbian plasticity struggle to learn. We show these representations are highly efficient, with sparse activations and low redundancy, similar to the ones learned by classic Hebbian/anti-Hebbian networks \cite{pehlevan2014hebbian}. Finally, we show that our learning rule naturally prevents catastrophic forgetting when several input data sets drawn from different distributions are presented to the network in succession. \

%Further investigation reveals that asynchronous learning is more robust to fast changing stimuli, high input size and high learning rates. 

\section{Lateral inhibition models}

Lateral inhibition was first proposed by Barlow in 1952 as a mechanism to encode sensory stimuli efficiently, the so-called redundancy reduction hypothesis \cite{barlow1961possible}. Initial implementations of such mechanisms can be dated back to Grossberg (1976) \cite{grossberg1976adaptive} and Rumelhart (1985) \cite{rumelhart1985feature} where they show that constant lateral inhibition between neurons leads to a competitive learning scheme in which a single neuron is active for a particular stimulus. Later Földiák (1990) pointed out that representations arising from competition are limited both in capacity and generalisation \cite{foldiak1990forming}. He proposed the first Hebbian/anti-Hebbian neural network model with plastic lateral inhibition. This was later shown to learn independent components of natural images (edges) \cite{falconbridge2006simple}, very similar to the features encoded by simple cells in the mammalian visual cortex \cite{hubel1962receptive}. More recently, models of the same flavor were mathematically derived from non-negative matrix factorization \cite{pehlevan2014hebbian} and similarity matching objectives \cite{pehlevan2017similarity}. \

\begin{figure}[!ht]
  \centering
  \includegraphics[width=\textwidth]{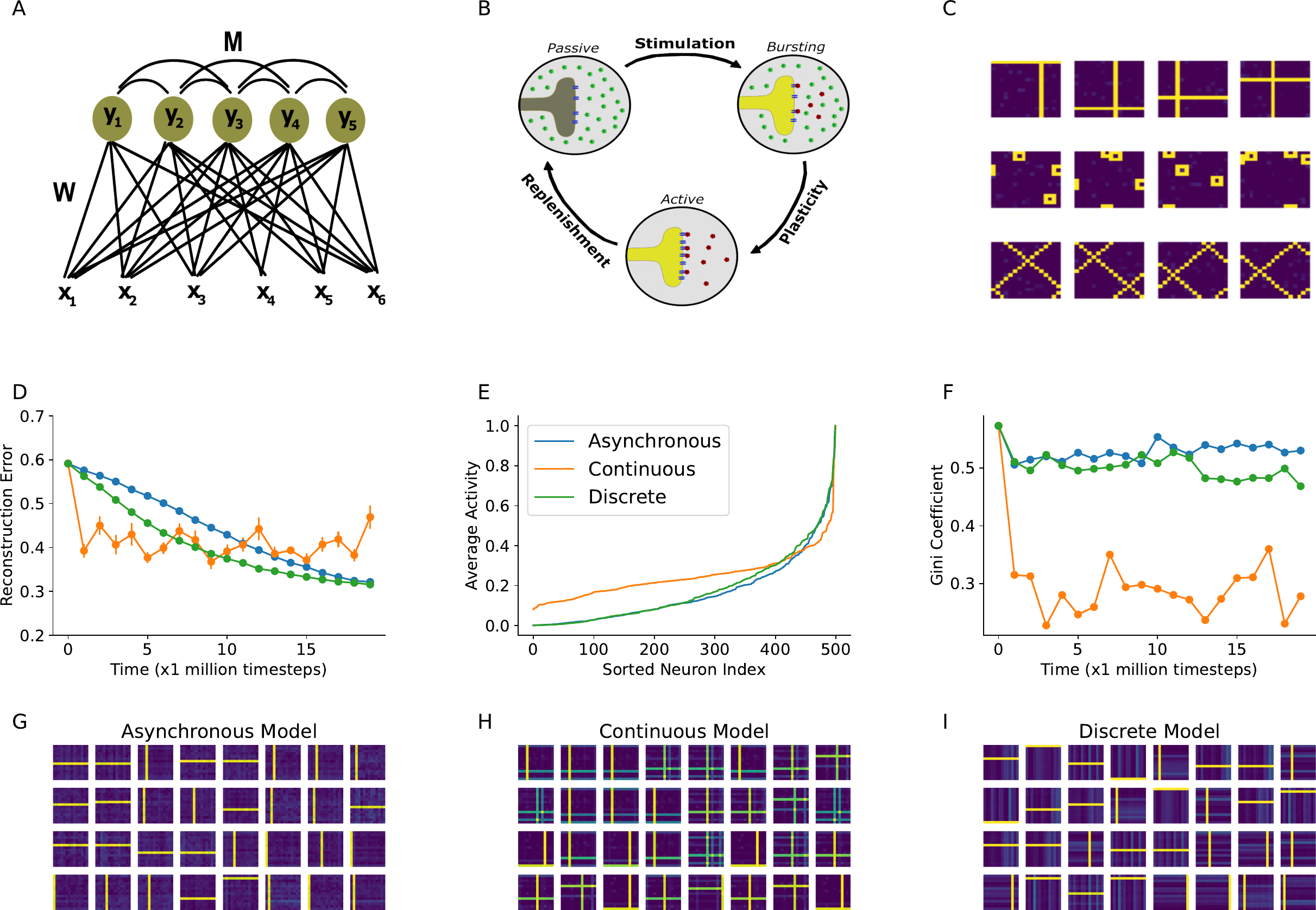}
  %\fbox{\rule[-.5cm]{0cm}{4cm} \rule[-.5cm]{4cm}{0cm}}
  \caption{{\bf Representation learning in Hebbian/anti-Hebbian networks.} (A) Diagram of the lateral inhibition model. Example shown has 5 neurons, each receives excitatory input from 6 external units ($x_{i}$) and inihibitory feedback from 5 recurrent units ($y_{i}$, including itself). (B) Schematic of the refractory mechanism leading to the asynchronous learning in a post-synpatic neuron. (C) Examples of three input stimulus sets shown to the model (top row: Földiák's bars). (D) Average reconstruction error for the three models during simulation with Földiák's bars (error bars are standard deviation over samples in each time bin). (E) Rank-ordered histogram of neural activity for the three models at the end of the simulation. (F) Gini coefficients of the activity distributions in D. A higher coefficient indicates higher lifetime sparseness. (G) Receptive fields of the 100 most active neurons of the asynchronuos network model. (H) As G, for the continuous network model. (I) As G, for the discrete network model.}
  \label{figure_intro}
\end{figure}

Figure \ref{figure_intro}A shows the general network model with feed-forward weights $\mathbf{W} \in \mathbb{R}^{n \times m}$ and recurrent weights $\mathbf{M} \in \mathbb{R}^{n \times n}$, where $m$ is the size of input $\mathbf{x}$ and $n$ in the number of neurons in the population $\mathbf{y}$. The dynamics of the activity is described by $\dot{\mathbf{y}} = [\mathbf{W} \mathbf{x} - \mathbf{M}\mathbf{y}]_{+}$ which we can simulate with random weights ($W$ and $M$) by showing a sample $\mathbf{x} = |x_{1} \quad \cdots \quad x_{n}|$ for a number of Euler steps $f$ (which we call the hold period as the sample $\mathbf{x}$ is kept constant throughout this period). If $M$ is a positive definite matrix, the system will eventually reach a stable state which we denote by $\mathbf{\hat{y}}$ (\cite{pehlevan2017similarity}). In the classical lateral inhibition model, once the network has reached the stable state, local Hebbian updates $\Delta w_{i,j} = x_jy_i - w_{i,j}$ on $W$ and $\Delta m_{i,j} = y_jy_i - m_{i,j}$ are applied on $M$. Note that due to the symmetric nature of the update, and the fact that $\mathbf{y}$ is always positive, $M$ will remain positive definite throughout the simulation, always guaranteeing a stable state as long as the hold period $f$ is long enough (usually we set $f=500$). Appendix A details other functional forms of local learning rules that have been proposed and the resulting receptive fields (see Appendix figure \ref{figure_learning_rules_rfs}). \

This model has been derived from the non-negative matrix factorization objective \cite{pehlevan2014hebbian} and here, we use it as a baseline, calling it the \emph{discrete model} (details in Appendix A3). A standard task this model can solve is pattern recognition in toy datasets. One can test this by training the model with a set of stimuli and analyzing the learned receptive fields of each neuron. Figure \ref{figure_intro}I shows some receptive fields this model learns when presented with Földiák's bars (Fig. \ref{figure_intro}C, top row). Such stimuli consist of crosses, which can be efficiently decomposed into stripes and bars. The discrete model learns precisely such feed-forward receptive fields, reproducing the results obtained by Földiák \cite{foldiak1990forming}. \

Central to the discrete model is the assumption that the neural dynamics is very fast and settles into an equilibrium point both before plasticity occurs and before the stimulus changes (i.e. $\mathbf{x}$ needs to be static until a stable state is reached). This assumption however is biologically implausible as neural plasticity is an ongoing process and stimulus changes may happen faster than neural dynamics. To investigate the importance of these assumption, we compare our model to a model which we call the {\em continuous model} (see Appendix B) where  plasticity is applied on par with the dynamics. In contrast to the discrete model, the continuous model does not learn decomposition into stripes and bars, but develops receptive fields containing crosses (Fig. \ref{figure_intro}H). Note that stimuli are still kept constant for a number of timesteps in the continuous model, however instead of a single plasticity update per sample (discrete model), we have $f$ plasticity updates per sample. \

We quantify the sparsity of the representation and observe that on average more neurons are active in the continuous model (Fig \ref{figure_intro}E). Activity sparseness can be quantified by the Gini coefficient (see Appendix E2), which tells us how efficient the representation is. This also relates to whether the model learned a suitable factorization or not (\cite{olshausen2004sparse}). We observe  a sparser representation in the discrete model  trend throughout the whole simulation (Fig \ref{figure_intro}F).  \

Furthermore, we compute the reconstruction error as a measure of encoding quality (for details see Appendix E1) and observe that the continuous model is more unstable in terms of representation quality (Fig. \ref{figure_intro}D). This may be due to a continuous drift of neuronal selectivity, which does not appear to be a feature of the discrete model (see Appendix figures \ref{figure_disc_evolution_crossbar} and \ref{figure_conti_evolution_crossbar}). We conclude that a sufficiently long hold period and waiting for the stable state of the dynamics is important as recurrent dynamics remove the statistical dependencies and redundancies in the neural activity. This, in turn, allows learning factorised representations. \

\begin{figure}[!ht]
  \centering
  \includegraphics[width=\textwidth]{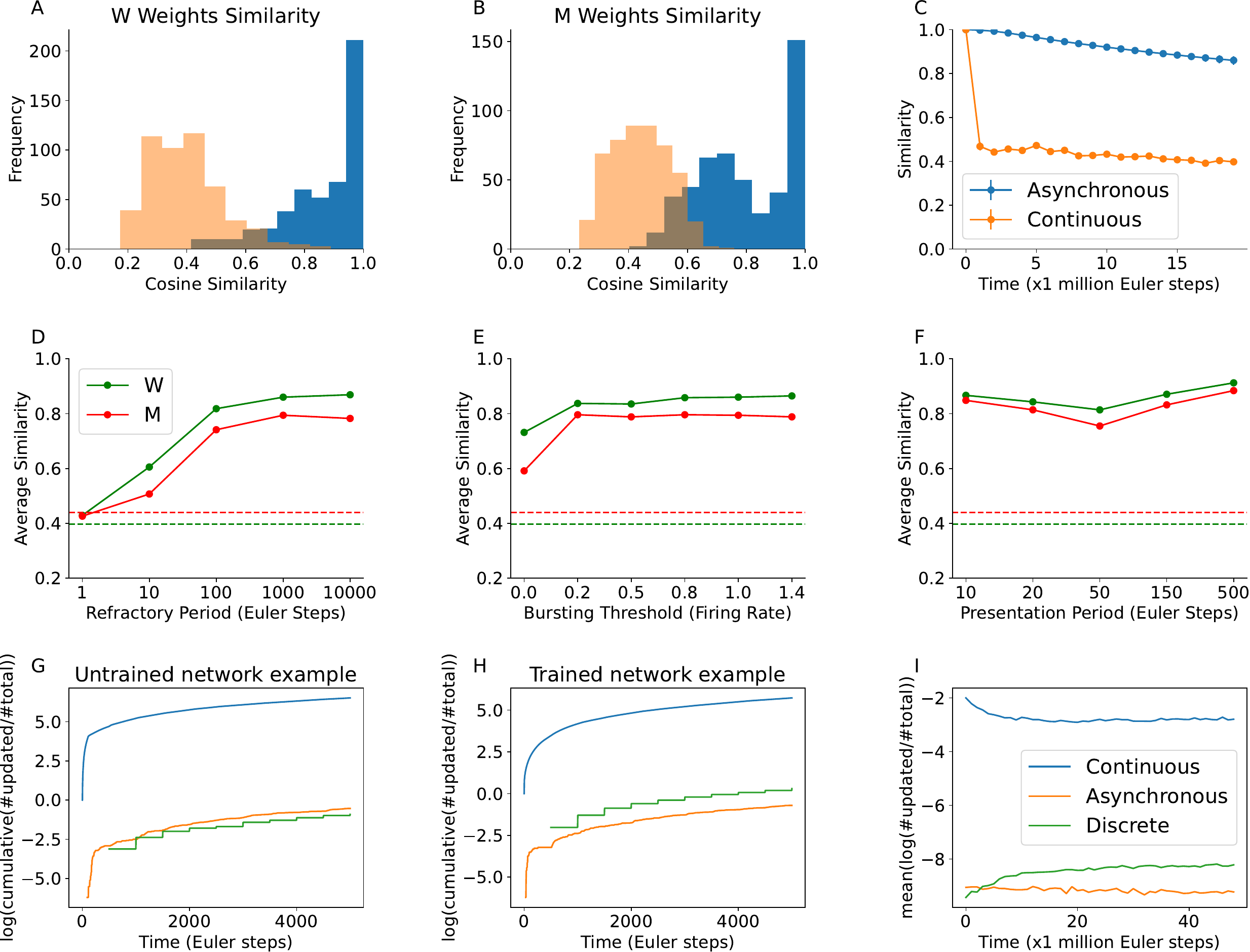}
  %\fbox{\rule[-.5cm]{0cm}{4cm} \rule[-.5cm]{4cm}{0cm}}
  \caption{{\bf The discrete and asynchronous models learn very similar representations.} (A) Histogram of cosine similarities of the feed-forward weight between the discrete model and the continuous (orange) and asynchronous (blue) model. (B) As A, for the recurrent weights. (C) Average cosine similarity of feed-forward weights, compared to the discrete model, as the simulation evolves (colors as in A). (D) Average cosine similarity of feed-forward (green) and recurrent (M) weights between discrete and asynchronous models for different refractory period durations in the asynchronous model. Dashed lines are similarities of the continuous model. (E) As D, for different bursting thresholds in the asynchronous model. (F) As D, for different presentation durations in the asynchronous model. (G) Short simulation window showing the number of synapses updated at each Euler step for untrained networks. (H) Same as G but for trained networks. (I) Average number of synaptic updates taken at uniform intervals throughout the whole simulation. }
  \label{figure_comparison}
\end{figure}

\section{Asynchronous Hebbian learning}

Here we introduce a new learning rule for lateral inhibition models. Instead of waiting for the stable state to apply an update (discrete model) or applying updates on par with the dynamics (continuous model), we propose to update neurons asynchronously: Each neuron updates its incoming weights when its activity has surpassed a threshold, and is unable to perform further updates for a \emph{refractory} period (see diagram in Fig. \ref{figure_intro}B; for equations, see Appendix C). Importantly, all updates in this \emph{asynchronous model} run in continuous time alongside the neural dynamics. The two crucial differences to the continuous model are: (1) only neurons whose receptive fields are sufficiently activated by an input will be updated, and (2) a selective neuron will not be drawn towards  other coinciding patterns because of the refractory period. \ 
%is this last sentence ok with the discretized version of the model?

We find the model learns a factorised representation (Fig. \ref{figure_intro}G) and maintains a similar level of sparseness compared to the discrete model (Fig. \ref{figure_intro}F). We also observe that both discrete and asynchronous models have very similar learning trajectories (Fig. \ref{figure_intro}D), reaching the same error at convergence. In contrast, the continuous model is more unstable and does not reach the same error. \

To assess the similarity between learning dynamics, we compare the learning trajectories of both asynchronous and continuous models with the discrete model. We initialize all models with the same weights and present the same stimulus sequence, and measure the cosine similarity of each neuron's incoming weights (see Appendix E3). Figures \ref{figure_comparison}A and \ref{figure_comparison}B show that after learning most neurons in the asynchronous model are practically identical (similarity 0.95-1.0) to the neurons in the discrete model, while the neurons in the continuous model diverge significantly. The divergence of the weights from the continuous model begins right at the start of training (Fig. \ref{figure_comparison}C), demonstrating that this network learns a qualitatively different representation. \

A key mechanism of our model is the refractoriness of plasticity which prevents a continuous update of the post-synaptic neuron's incoming weights while it is bursting. Figure \ref{figure_comparison}D shows that refrectoriness is quite important for the asynchronous model to approximate the learning trajectory of the discrete model, as non-existent (1 step) or small (10 steps) refractoriness lead to poor average weight similarity. Interestingly, this refractory period has also been observed in \emph{in vitro} experiments \cite{flores2024synapse}. Also note that without a refractory period this model will learn to a non-factorised, winner-take-all representation similar to the one learned by the continuous model (see Appendix figures \ref{figure_async_crossbar_refractoriness_rfs} and \ref{figure_async_crossbar_refractoriness_histograms}). Varying the threshold for bursting does not affect the learning much unless we set it to zero, in which case the network seems to diverge from the discrete version (Fig. \ref{figure_comparison}E). Varying the hold period (i.e. the number of iterations the stimuli is held for the network to reach a stable state) does affect the learning trajectory (Fig. \ref{figure_comparison}F) which is interesting since the standard version of discrete network (which uses the same learning rule as our model - Hebbian) stops learning as the hold period goes below 150 (see Appendix figures \ref{figure_learning_rules_rfs}). \

We further explore how learning differs on these models by counting the number of synapses that are updated (i.e. have gradient entry different from 0). Figure \ref{figure_comparison}G and \ref{figure_comparison}H show the number of synapses updated at each Euler step during a small simulation window. As expected, the discrete network has a stair-case like shape since it only updates once every 500 steps (i.e. the hold period for this simulation). It is interesting to note that the asynchronous network follows a very similar trajectory to the discrete network for a random untrained network (Fig. \ref{figure_comparison}G). However, as we train the networks, the discrete model seems to increase the number of updates while the asynchronous model slightly decreases them (Fig. \ref{figure_comparison}I). \

\begin{figure}[!ht]
  \centering
  \includegraphics[width=\textwidth]{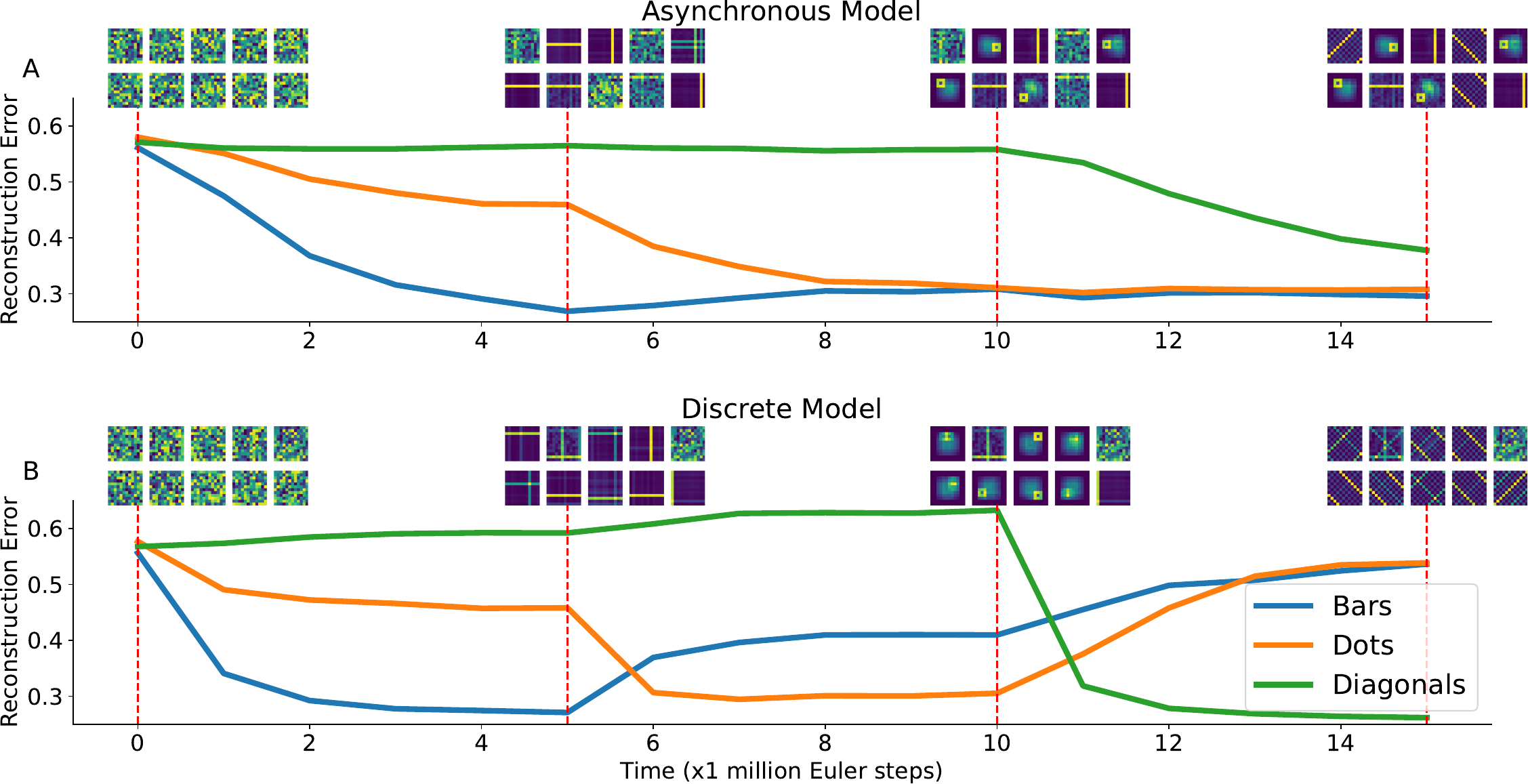}
  %\fbox{\rule[-.5cm]{0cm}{4cm} \rule[-.5cm]{4cm}{0cm}}
  \caption{{\bf The asynchronous model prevents catastrophic forgetting.} Three differnt sets of stimuli (see Fig. 1C) were shown in succession during ongoing plasticity (red bars indiacte when the stimulus set changed). The reconstruction error was computed for all stimuli and is shown for the  asynchronous model (A) and  the discrete model (B). Once a stimulus set has been learned by the asynchronous model, the low error persists when the next set is introduced, while it increases again in the discrete model. Insets show the same, randomly selected, feed-forward weights, throughout each simulation.}
  \label{figure_forgetting}
\end{figure}

\section{Continual learning}

Biological neuronal networks have the ability to maintain task performance, and learn new tasks without forgetting previous information. Artificial neural networks, in contrast, suffer from catastrophic forgetting, where a task previously learned is forgotten when a new task is learned \cite{french1999catastrophic}. Here we find that the asynchronous rule naturally prevents catastrophic forgetting as long as the network capacity is sufficient to represent all relevant factors. \

To show this, we present three different sets of stimuli (Fig. \ref{figure_intro}C) in sequence in three different phases, and continuously test the ability of the model to reconstruct all three stimuli sets (Fig. \ref{figure_forgetting}, vertical red dashed lines; insets show a selected subset of receptive fields at the points when stimuli distribution is changed). We observe that the asynchronous model retains the receptive fields learned from previous phases, and therefore can maintain a low reconstruction error for all three stimuli at the end of the simulation. In contrast, the discrete model constantly adapts all receptive fields to the new distribution, hence forgetting previous patterns and generating high errors at the end of the simulation for both stimuli 1 and 2 (Fig. \ref{figure_forgetting}B). In the asynchronous model, unstructured and redundant receptive fields are primarily used to encode new data (see insets in Fig. \ref{figure_forgetting}A; see more examples in the Appendix figures \ref{figure_async_evolution_mixed_stimuli} and \ref{figure_disc_evolution_mixed_stimuli}). Catastrophic forgetting is avoided in the asynchronous model because the plasticity threshold prevents plasticity in weakly tuned neurons and, acting in concert with lateral inhibition, reducing the likelihood of plasticity in weakly activated neurons that are tuned to previously learned stimuli. \

%Comment on how the stimuli are stored, i.e. non-overlapping groups of RFs.

\section{Discussion}

In this work, we propose a biologically plausible mechanism to learn efficient factorised representations of inputs in lateral inhibition models with time-continuous plasticity. We show that it approximates the learning of classic Hebbian/anti-Hebbian networks derived from the non-negative matrix factorization. We also show that the same mechanism effectively prevents catastrophic forgetting. The emerging representations are causal models of the input ensemble, and they resemble those of blind source separation algorithms such as ICA \cite{bell1995information}. Interestingly, the predictive coding literature contains a range of models based on expectation-maximization where dynamics are used to reach a stable state before a plasticity update is applied. This includes biological implementations of the back-propagation algorithm such as equilibrium propagation \cite{scellier2017equilibrium}. Applying our asynchronous mechanism to learn energy-based models could be a promising avenue for future work. \

In contrast to the group of "discrete" models which require settling of recurrent dynamics, the asynchronous model is biologically more plausible in several ways: Plasticity events are more likely during bursts of spikes \cite{paulsen2000natural,froemke2006contribution,inglebert2020synaptic}, and refractoriness of plasticity has been reported in experiments \cite{kramar2012synaptic,flores2024synapse}. Furthermore, the asynchronous model produces fewer plasticity events than a continuous model while still learning at a similar rate. It has been suggested that plasticity events are metabolically costly and it may be a biological objective to limit them in the brain \cite{pache2023energetically}. The asynchronous mechanism is a candidate model implementing this constraint. \

One important question is whether the asynchronous model shows a similar behaviour in spiking networks, as we propose it as a biological mechanism. Spiking networks with lateral inhibition and Hebbian plasticity have been shown to approximate ICA-like representations in the presence of a homeostatic mechanism which maintains an appropriate target activity level and enables stable competitive learning \cite{triesch2004synergies, savin2010independent}. While in these studies the weights are updated continuously, sparse spiking may restrict plasticity events in a similar way to refractoriness in our model. A systematic comparison between spiking and rate-based models will therefore be of interest, and holds promise as it can lead to a normative understanding of neural plasticity. \

\bibliography{references}

%%%%%%%%%%%%%%%%%%%%%%%%%%%%%%%%%%%%%%%%%%%%%%%%%%%%%%%%%%%%
\appendix
\renewcommand\thefigure{\thesection.\arabic{figure}}
\setcounter{figure}{0}

\section{Local learning rules}

Hebbian/anti-Hebbian networks have been extensively studied and analysed, multiple functional forms have been proposed in the literature. Here we briefly review different functional forms of local learning rules and show what they learn when presented with our set of stimuli. Following the classic learning procedure of Hebbian/anti-Hebbian networks \cite{foldiak1990forming, falconbridge2006simple, pehlevan2014hebbian}, we present a stimulus $\mathbf{x} \in \mathbb{R}^{m}$ to the network and let the dynamics settle into a stable state for a fixed number of Euler iterations (see Appendix D). From this we obtain the stable state representation $\mathbf{\hat{y}} = \left| \hat{y}_{1} \quad \hat{y}_{2} \quad \cdots \quad \hat{y}_{n} \right|$ for stimulus $\mathbf{x} = \left| x_{1} \quad x_{2} \quad \cdots \quad x_{m} \right|$. We tested also smaller presentation periods as shown in figure A1 and A2. For each update we keep all weights positive in order to keep feed-forward weights strictly excitatory and recurrent weights strictly inhibitory. In the following subsections we describe the different functional forms that different works have proposed. \

\begin{figure}[!ht]
  \centering
  \includegraphics[width=\textwidth]{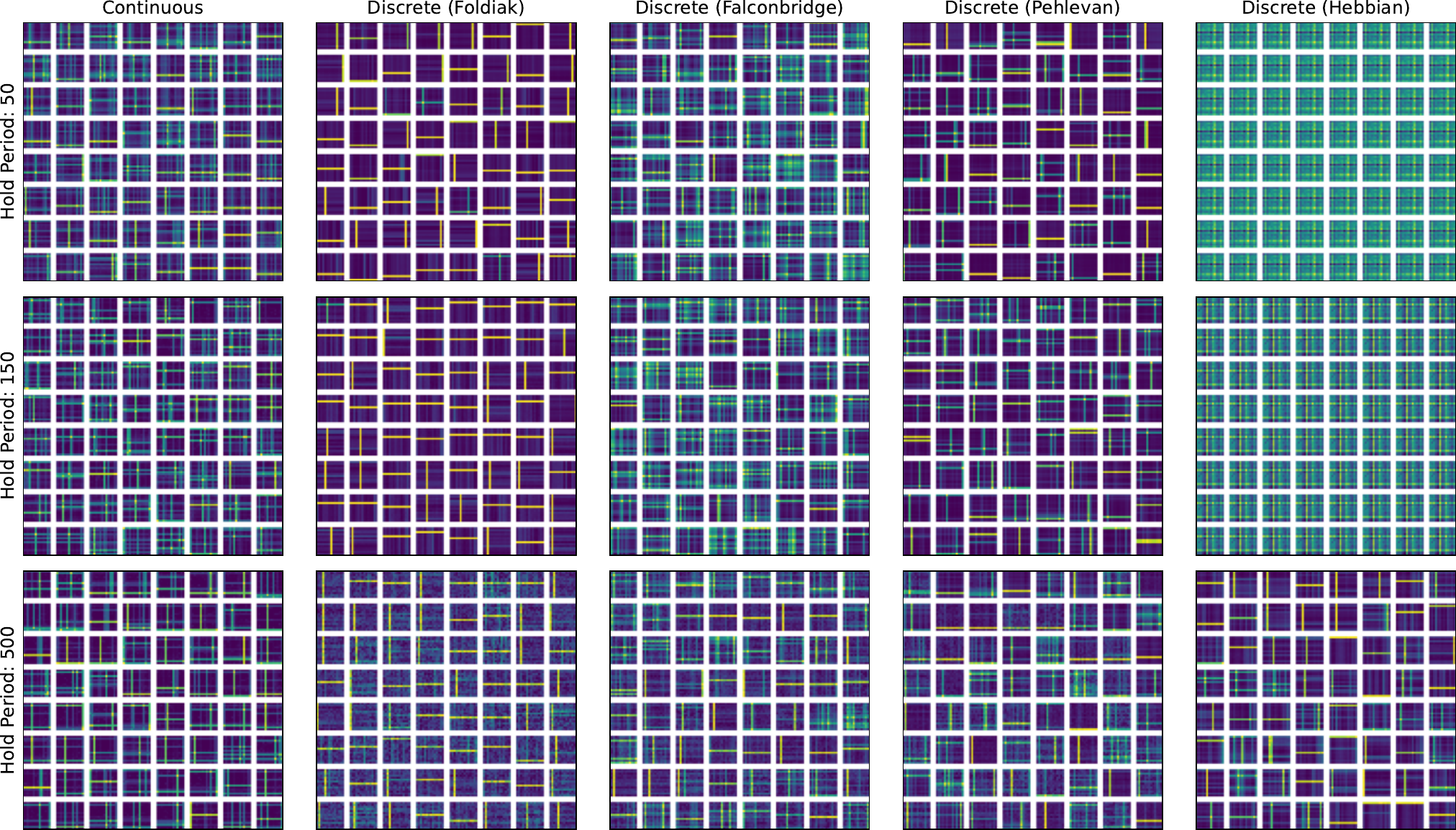}
  %\fbox{\rule[-.5cm]{0cm}{4cm} \rule[-.5cm]{4cm}{0cm}}
  \caption{Receptive fields of the top 100 most active neurons for the continuous model and different versions of the discrete model. Models were simulated with different presentation (holding) times, quantified by Euler steps (rows)}
  \label{figure_learning_rules_rfs}
\end{figure}

\begin{figure}[!ht]
  \centering
  \includegraphics[width=\textwidth]{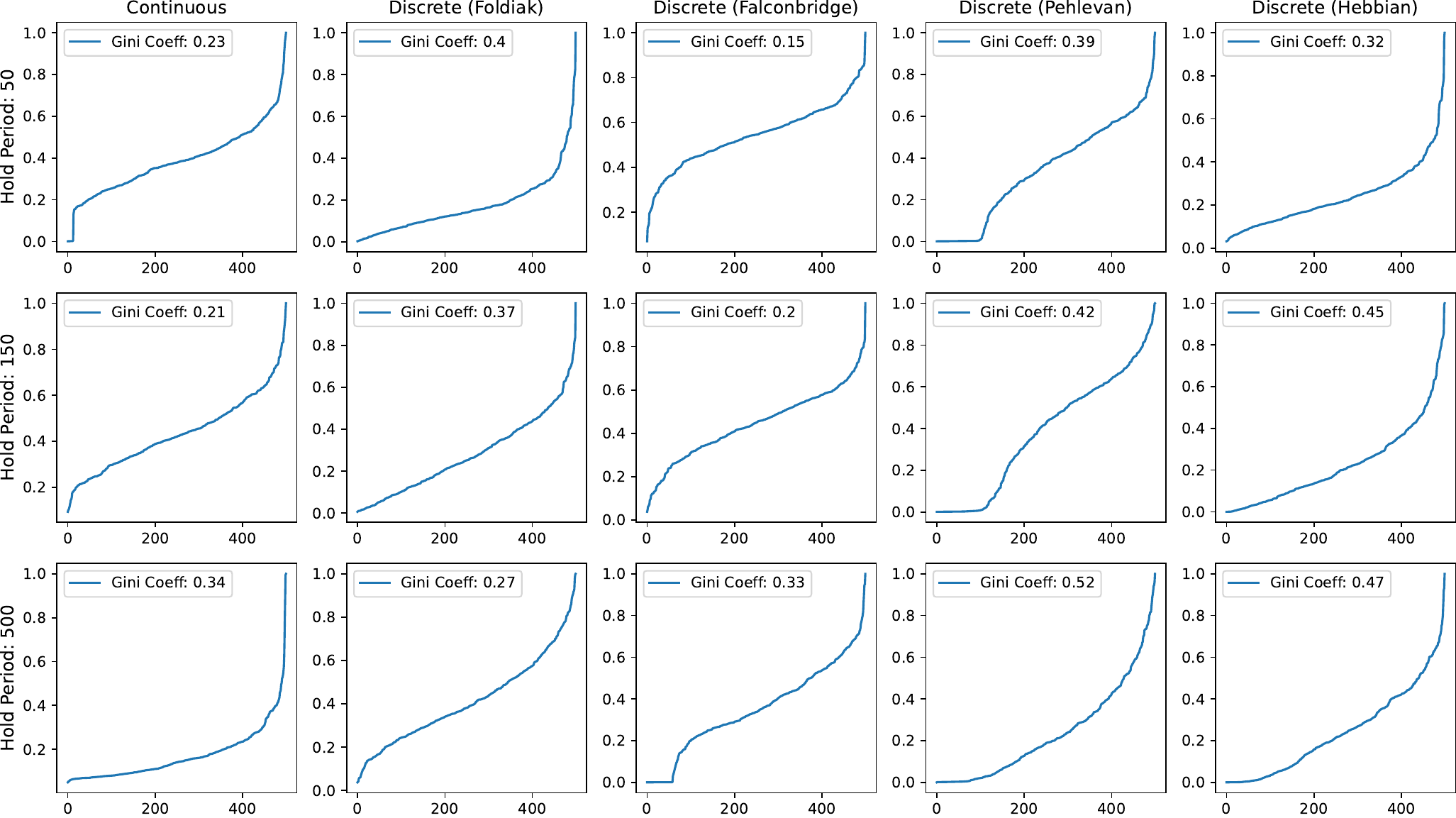}
  %\fbox{\rule[-.5cm]{0cm}{4cm} \rule[-.5cm]{4cm}{0cm}}
  \caption{Histogram of average neural activity for the continuous model and different versions of the discrete model. Models were simulated with different presentation (holding) times, quantified by Euler steps (rows). For each plot, x-axis is the sorted neuron index and the y axis in the normalized average activity. }
  \label{figure_learning_rules_histograms}
\end{figure}

\subsection{Földiák network}

The classic Földiák network \cite{foldiak1990forming} includes a bias in the dynamics, sometimes called the sensitivity, however, we only consider the update rules of feed-forward and recurrent weights and neglect the sensitivity term so as to compare identical network models. The feed-forward update is the following:

\begin{equation}
    \Delta w_{i,j} = \eta (\hat{y}_{i}x_{j} - \hat{y}_{i}w_{i,j})
\end{equation}

Here, $w_{i,j}$ is the entry $i,j$ of the feed-forward matrix $\mathbf{W}$ and $\eta$ is a learning rate (usually $0.01$ for our simulations). The recurrent weight update is:

\begin{equation}
    \Delta m_{i,j} = \eta (\hat{y}_{i}\hat{y}_{j} - p^2)
\end{equation}

Again, $\eta$ is a learning rate (usually $0.01$ for our simulations), $m_{i,j}$ is the entry $i,j$ of the recurrent matrix $\mathbf{M}$ and $p$ is a small positive constant ($p << 1$).

\subsection{Falconbridge network}

Falconbridge et al \cite{falconbridge2006simple} proposed a very similar network to Földiák and suggested it learns the independent components of natural images \cite{bell1997independent}, following a similar learning trajectory to the sparse coding network \cite{olshausen2004sparse}. The only difference from Földiák's model is the feed-forward weight update which is the same as the one proposed by Oja \cite{oja1982simplified}:

\begin{equation}
    \Delta w_{i,j} = \eta (\hat{y}_{i}x_{j} - \hat{y}_{i}^{2}w_{i,j})
\end{equation}

\subsection{Pehlevan network}

More recently, Pehlevan and Chklovskii derived the network updates from the non-negative matrix factorization objective \cite{pehlevan2014hebbian}. They obtained identical rules for both the feed-forward and recurrent weights, with the addition of a dynamics learning rate:

\begin{equation}
    \Delta w_{i,j} = \eta_{i} (\hat{y}_{i}x_{j} - \hat{y}_{i}^{2}w_{i,j})
\end{equation}

\begin{equation}
    \Delta m_{i,j} = \eta_{i} (\hat{y}_{i}\hat{y}_{j} - \hat{y}_{i}^{2}m_{i,j})
\end{equation}

\begin{equation}
    \Delta \eta_{i} = \hat{y}_{i}^{2}
\end{equation}

Note that post-synaptic neurons evolve their own distinct learning rates based on their activity, which resembles the way in which we restrict updates on post-synaptic neurons for our asynchronous model. \

\subsection{Hebbian network}

A simpler version of the Hebbian/anti-Hebbian network has been proposed to explain certain properties of receptive fields in the cortex such as representational drift \cite{qin2023coordinated}. Such version can be formulated as follows:

\begin{equation}
    \Delta w_{i,j} = \eta (\hat{y}_{i}x_{j} - w_{i,j})
\end{equation}

\begin{equation}
    \Delta m_{i,j} = \eta (\hat{y}_{i}\hat{y}_{j} - m_{i,j})
\end{equation}

\section{Continuous model}

The implementation of the continuous model is straight-forward, instead of updating the weights with the stable state activity $\mathbf{\hat{y}}$, we apply an update every time we apply an Euler step in the simulation. We present an input $\mathbf{x}$ to the network for $f$ Euler steps and for each step $t$ we compute the activity vector $\mathbf{y}_{t} = = \left| \hat{y}_{1} \quad \hat{y}_{2} \quad \cdots \quad \hat{y}_{n} \right|$ which is then used to perform the weight updates of both feed-forward and recurrent weights using Oja's rule and a fixed learning rate of $\eta = 0.001$:

\begin{equation}
    \Delta w_{i,j} = \eta (y_{t,i}x_{j} - y_{t,i}^{2}w_{i,j})
\end{equation}

\begin{equation}
    \Delta m_{i,j} = \eta (y_{t,i}y_{j} - y_{t,i}^{2}m_{i,j})
\end{equation}

Note that for each data point $\mathbf{x}$ we perform $f$ updates in both the weight and the dynamics. This is in contrast to the discrete model, where only a single weight update is performed for each data stimulus. \

\section{Asynchronous learning mechanism}

In this paper we introduce a mechanism that allows each neuron in the network to learn independently of each other without having to wait for the whole network to reach a stable state. This mechanism consists of updating the weights of a single post-synaptic neuron only when the neuron has a burst of activity. Following the burst and the update, this neuron cannot modify its weights for a small period of time (refractory period). \\

Let $y_{i}$ be the activity of a post-synaptic neuron $i$ and $x_{j}$ be the activity of the pre-synaptic neuron. We introduce a new variable $\beta_{i} \in {0,1}$ which gates the continuous plasticity update:

\begin{equation}
    \Delta w_{i,j} = \beta_{i}(y_{i}x_{j} - w_{i,j})
    \label{eq:oja_rule}
\end{equation}

The variable $\beta_{i}$ is updated as follows:

  \[
        \beta_{i}\leftarrow\left\{
                \begin{array}{ll}
                  1 \textnormal{ if } y_{i} > r_{b} \textnormal{ and } c_{i} > r_{r}\\
                  0 \textnormal{ else}
                \end{array}
              \right.
  \]

The constant $r_{b}$ is the activity threshold above which a neuron is considered bursting and plasticity can occur. $r_{r}$ is the length of the refractory period defined here as the number of Euler steps that have to elapse until the neuron can update again.

$c_{i}$ is updated as a simple counter:

  \[
        c_{i}\leftarrow\left\{
                \begin{array}{ll}
                  0 \textnormal{ if } y_{i} > r_{b} \textnormal{ and } c_{i} > r_{r}\\
                  c_{i}+1 \textnormal{ else}
                \end{array}
              \right.
  \]

\begin{figure}[!ht]
  \centering
  \includegraphics[width=\textwidth]{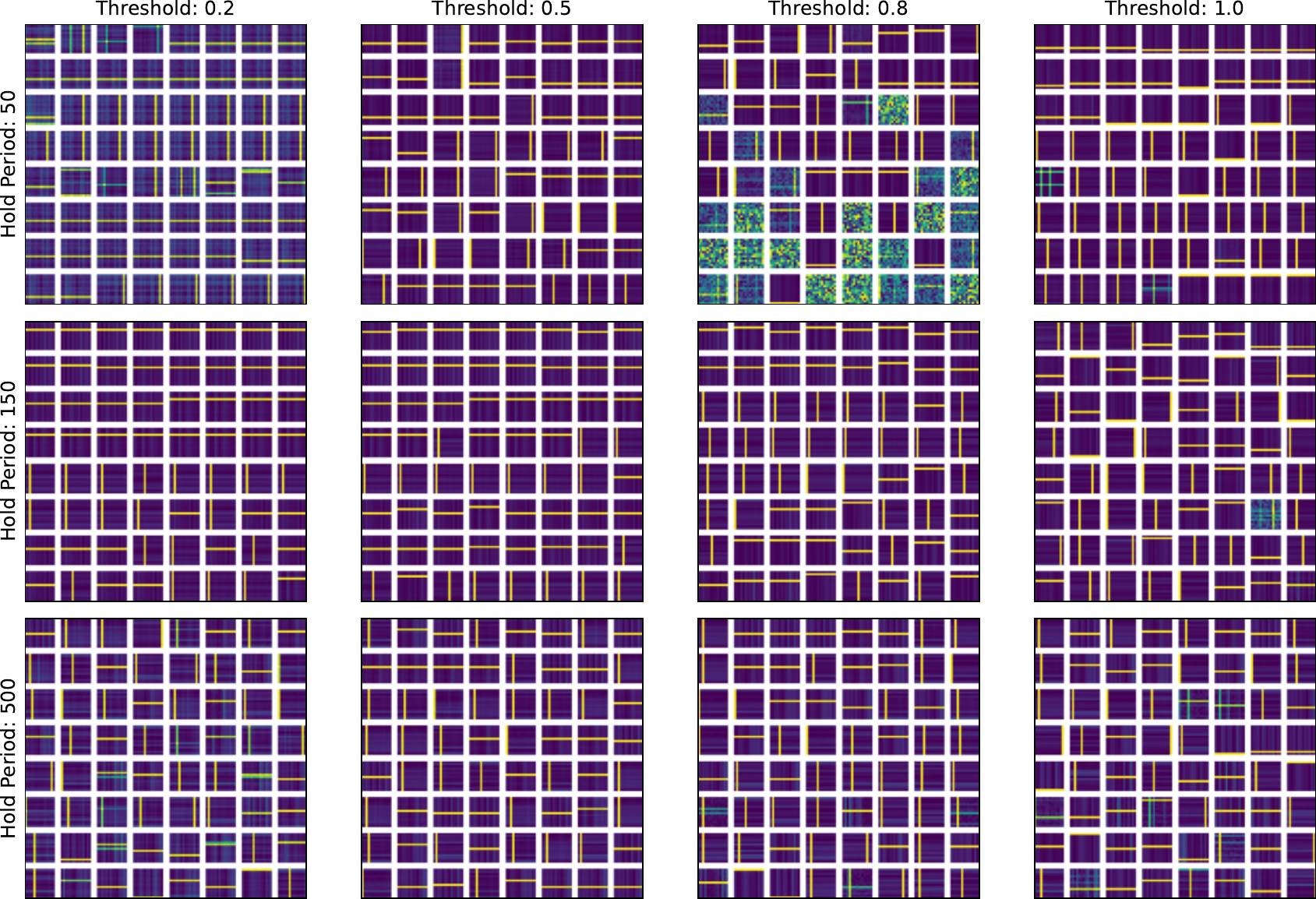}
  %\fbox{\rule[-.5cm]{0cm}{4cm} \rule[-.5cm]{4cm}{0cm}}
  \caption{Receptive fields of top 100 most active neurons for different bursting thresholds of the asynchronous model. Models were simulated with different presentation (holding) times, quantified by Euler steps (rows)}
  \label{figure_async_crossbar_thresholds_rfs}
\end{figure}

\begin{figure}[!ht]
  \centering
  \includegraphics[width=\textwidth]{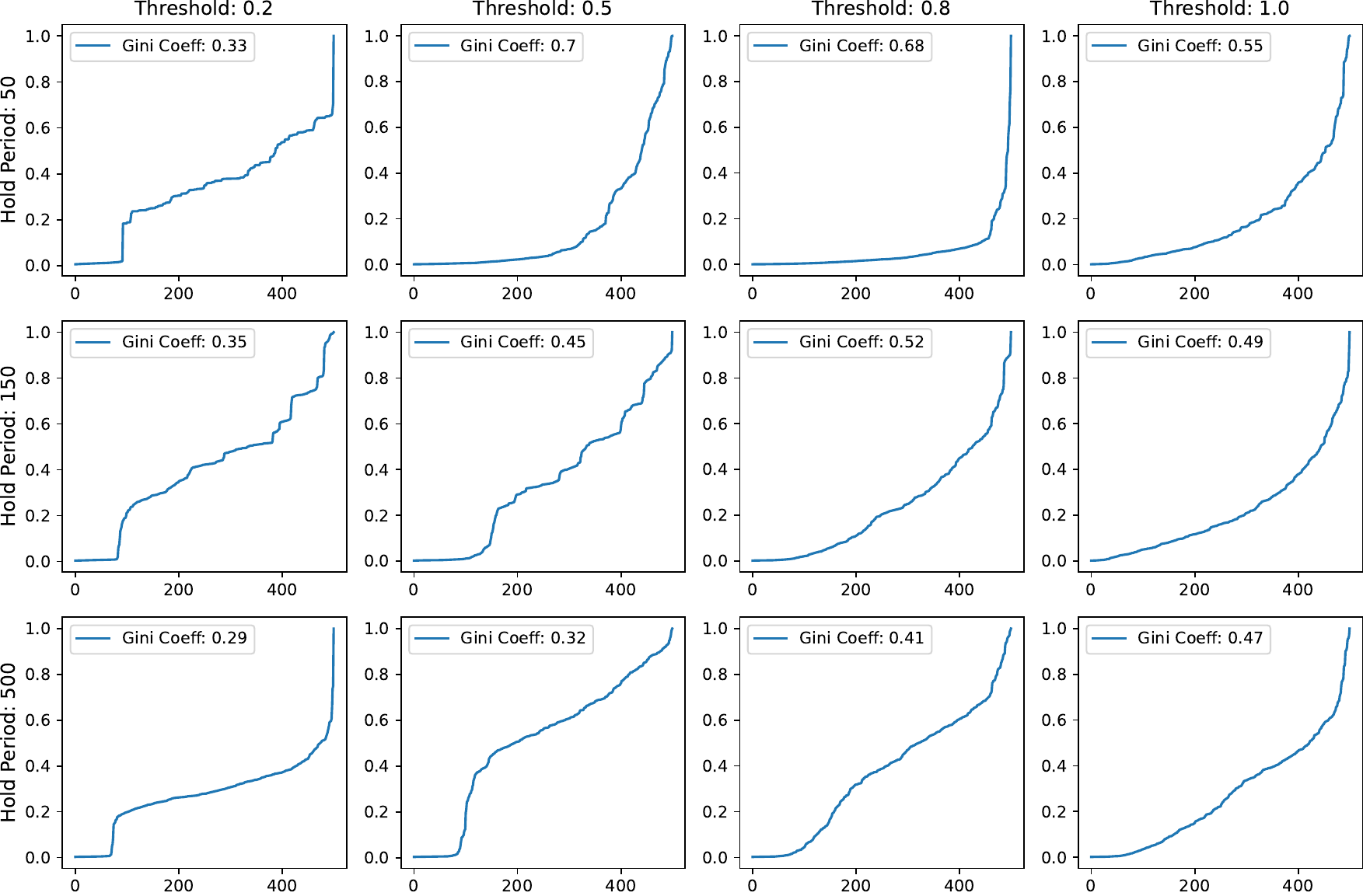}
  %\fbox{\rule[-.5cm]{0cm}{4cm} \rule[-.5cm]{4cm}{0cm}}
  \caption{Histogram of average neural activity for different bursting thresholds of the asynchronous model. Models were simulated with different presentation (holding) times, quantified by Euler steps (rows). For each plot, x-axis is the sorted neuron index and the y axis in the normalized average activity.}
  \label{figure_async_crossbar_thresholds_histograms}
\end{figure}

\begin{figure}[!ht]
  \centering
  \includegraphics[width=\textwidth]{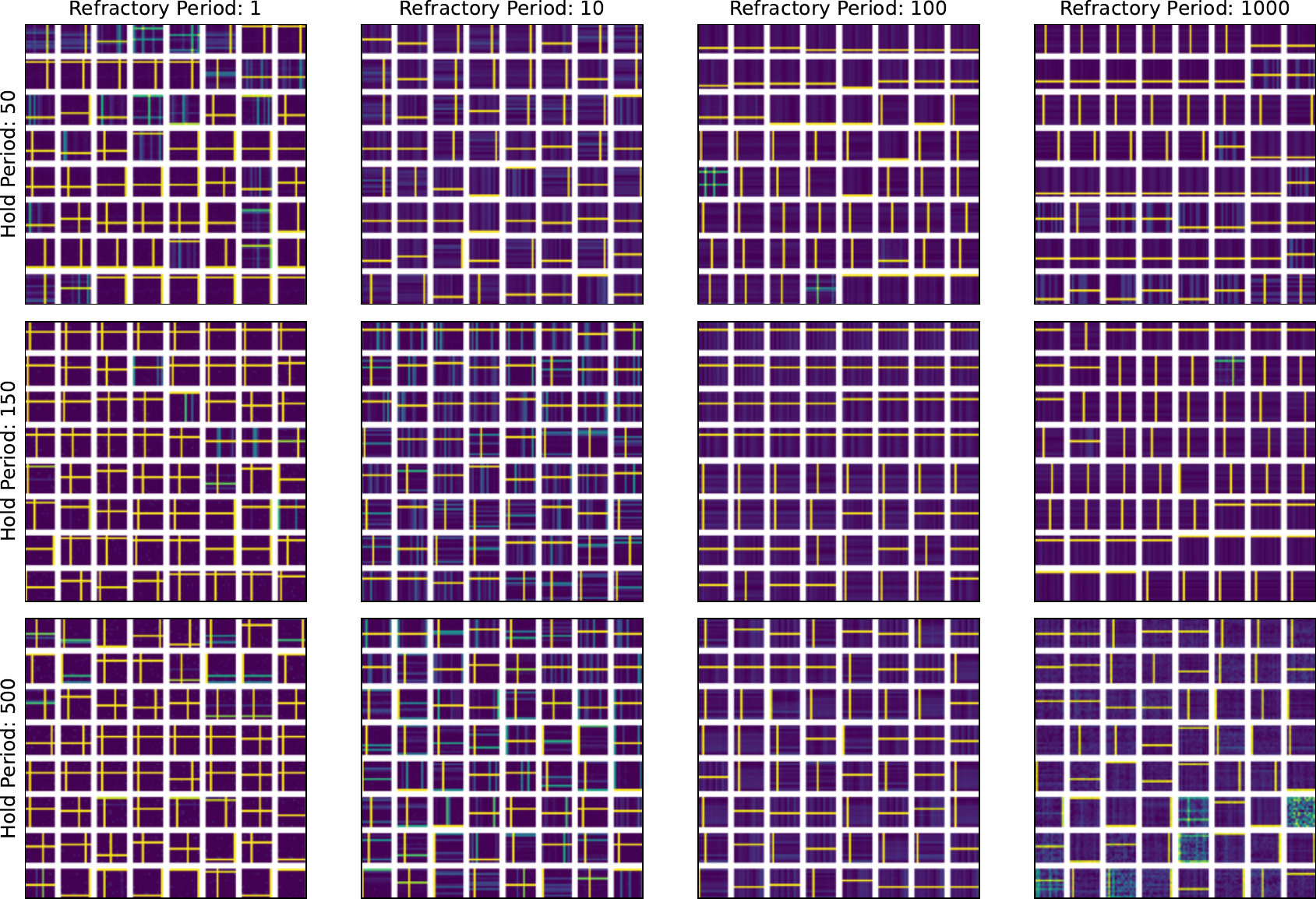}
  %\fbox{\rule[-.5cm]{0cm}{4cm} \rule[-.5cm]{4cm}{0cm}}
  \caption{Receptive fields of top 100 most active neurons for different refractory periods of the asynchronous model. Models were simulated with different presentation (holding) times, quantified by Euler steps (rows)}
  \label{figure_async_crossbar_refractoriness_rfs}
\end{figure}

\begin{figure}[!ht]
  \centering
  \includegraphics[width=\textwidth]{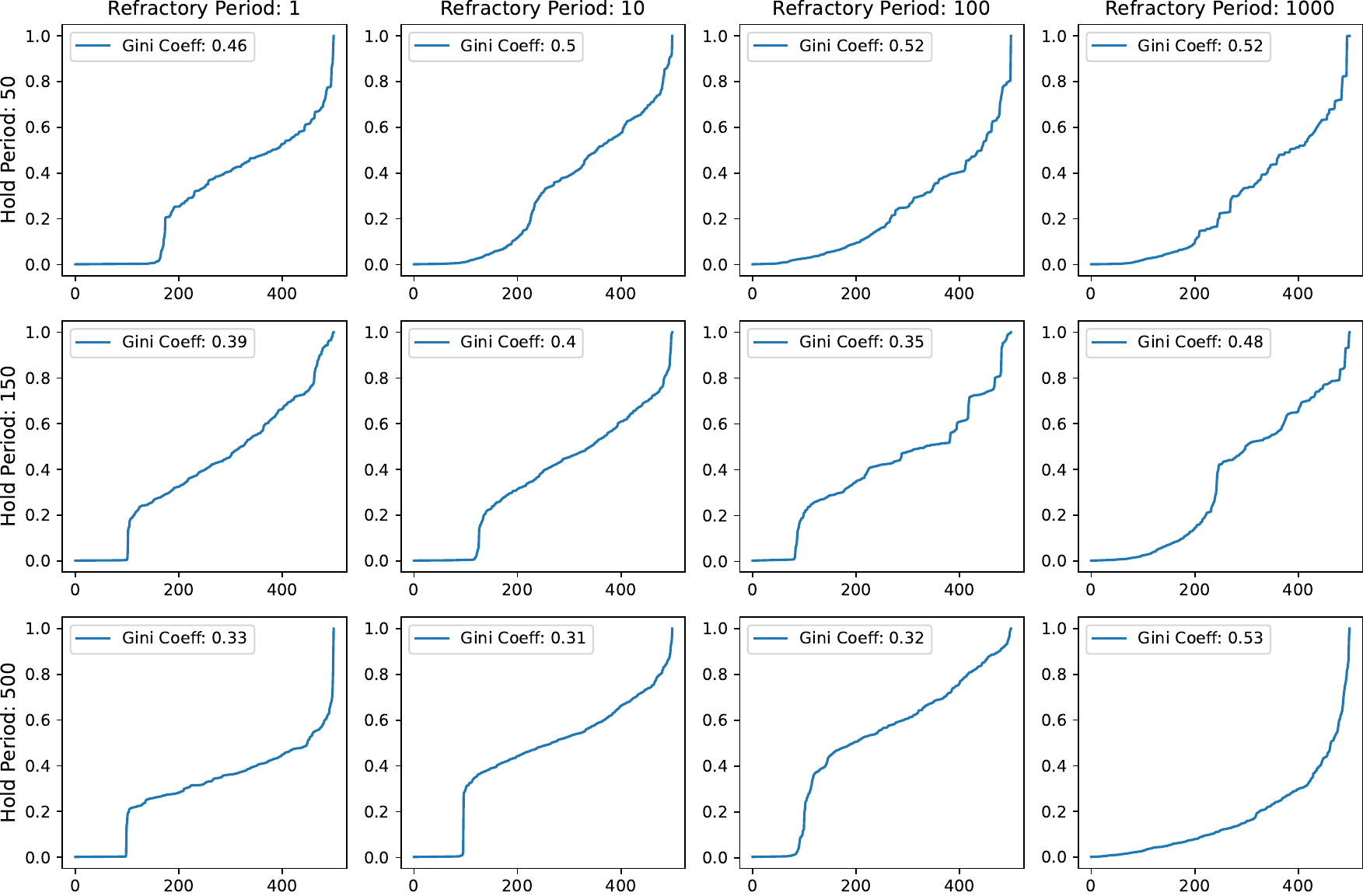}
  %\fbox{\rule[-.5cm]{0cm}{4cm} \rule[-.5cm]{4cm}{0cm}}
  \caption{Histogram of average neural activity for different refractory periods of the asynchronous model. Models were simulated with different presentation (holding) times, quantified by Euler steps (rows). For each plot, x-axis is the sorted neuron index and the y axis in the normalized average activity.}
  \label{figure_async_crossbar_refractoriness_histograms}
\end{figure}

\section{Simulation details}

For all models we set the number of neurons $n = 500$ and set the input size $m=14\times14$ (gray-scale pixels, s.t. $x_{i} \in [0,1]$). Learning rates for the discrete model (except for the Pehlevan network) were $\eta=0.01$ and for the continuous model we set $\eta=0.001$. All three models had the same dynamics which were simulated via Euler's method with a step size of $0.01$ and we tested three different hold periods (i.e. number of Euler iterations were input was kept constant) - $50$, $150$ and $500$. We initialized the weights as random vectors were each entry was drawn from a Guassian distribution with mean $0$ and variance $1$. All the weights were kept positive throughout the entire simulations. \

Python was utilized to implement the models and evaluation metrics. The libraries utilized were \emph{numpy}, \emph{matplotlib} and \emph{scipy}. All simulations can be run on a regular desktop for a few hours and the code to reproduce the experiments is open-source \footnote{\url{https://github.com/henri-edinb/async\_learning}}. Servers with many CPUs were also used to accelerate experiments but are not necessary as $4$GB RAM should be enough to run networks of this size. \

\section{Evaluation Methods}

A straight-forward approach to verify whether a single-layer network has learned a good representation is to visualise the receptive fields. From Földiák's bars, one would expect a fully competitive network to learn receptive fields that look like the input (i.e. crosses). If the network successfully learns a factorised representation then we expect to see stripes and bars \cite{foldiak1990forming}. \

For a quantitative analysis of the behaviour, we obtain the reconstruction error and sparseness of the average activity measured via the Gini coefficient. Below we describe both these methods and also explain how we compared the learning trajectories of the models. \

\subsection{Reconstruction error}

To analyse whether the network has learned the patterns in the input, we reconstruct the input from the latent representation at the stable state by utilizing the transpose of the feed-forward weights $\mathbf{W}^{T}$. In all three models (asynchronous, continuous and discrete), we present a sequence of stimuli $\{\mathbf{x}_{i}\}_{i=0}^{s}$, hold the input $\mathbf{x}_{i}$ for $f$ Euler steps with size $0.01$ and compute the reconstruction $\mathbf{\overline{x}}_{i,t}$ at every Euler step $t$. We then measure the reconstruction error and average it over the the whole presentation:

\begin{equation}
\sum_{i=0}^{s}\sum_{t=0}^{f} \frac{1 - sim(\mathbf{x}_{i}, \mathbf{\overline{x}}_{i,t})}{s*f}
\end{equation}

Throughout our simulations, we set $s=60$ and $f=150$ which allows for a fair comparison between models. The similarity measure is the cosine similarity:

\begin{equation}
sim(\mathbf{a}, \mathbf{b}) = \frac{\mathbf{a}^{T}\mathbf{b}}{\left\| \mathbf{a} \right\| * \left\| \mathbf{b} \right\|}
\label{eq:similarity}
\end{equation}

\subsection{Gini coefficient}

To measure the sparness of activity of the network we first take the average activity over a time window similar to the reconstruction error. We present a sequence of stimuli $\{\mathbf{x}_{i}\}_{i=0}^{s}$, hold the input $\mathbf{x}_{i}$ for $f$ Euler steps with size $0.01$ and obtain the firing rate at each Euler step $\mathbf{y}_{i,t}$. We then average the activity over the test window obtaining a vector $\widetilde{\mathbf{y}}$ with the average activity:

\begin{equation}
\widetilde{\mathbf{y}} = \sum_{i=0}^{s}\sum_{t=0}^{f} \mathbf{y}_{i,t}
\end{equation}

With this we can compute the Gini coefficient, which yields a quantitative measure of the sparseness of the activity:

\begin{equation}
G(\mathbf{y}) = \frac{\sum_{i=0}^{n}\sum_{j=0}^{n}\left| y_{i} - y_{j} \right|}{2*n*\sum_{i=0}^{n}y_{i}}
\end{equation}

\subsection{Weight comparison}

Since all models have the same weight structure, we can compare their learning trajectory by measuring the cosine similarity between the incoming weights of each neuron. Let a neuron $i$ from model $A$ (resp. $B$) have feed-forward weights $\mathbf{W}_{A}$ (resp. $\mathbf{W}_{B}$) and recurrent weights $\mathbf{M}_{A}$ (resp. $\mathbf{M}_{B}$). Also let the $i$th row of the feed-forward matrix $\mathbf{W}_{A}$ (resp. $\mathbf{W}_{B}$) be denoted by the vector $\mathbf{w}_{A,i}$ (resp. $\mathbf{w}_{B,i}$) and the $i$th row of the recurrent matrix $\mathbf{M}_{A}$ (resp. $\mathbf{M}_{B}$) be denoted by the vector $\mathbf{m}_{A,i}$ (resp. $\mathbf{m}_{B,i}$). Note these vectors we defined are the incoming weights of post-synaptic neuron $i$. To compare the feed-forward (resp. recurrent) weights of model $A$ and $B$ we measure the cosine similarity between $\mathbf{w}_{A,i}$ and $\mathbf{w}_{B,i}$ (resp. $\mathbf{m}_{A,i}$ and $\mathbf{m}_{B,i}$) using equation \ref{eq:similarity}. We do this over all neurons in the models and bin the results to obtain the histograms plotted in Figures 2A and 2B. To obtain the rest of the figures, we compute the mean and variance of the results. \

\section{Receptive fields temporal evolution}
%%%%%%%%%%%%%%%%%%%%%%%%%%%%%%%%%%%%%%%%%%%%%%%%%%%%%%%%%%%%
The rest of the appendix show figures with the evolution of the receptive fields for different models.
\

\begin{figure}[!ht]
  \centering
  \includegraphics[width=\textwidth]{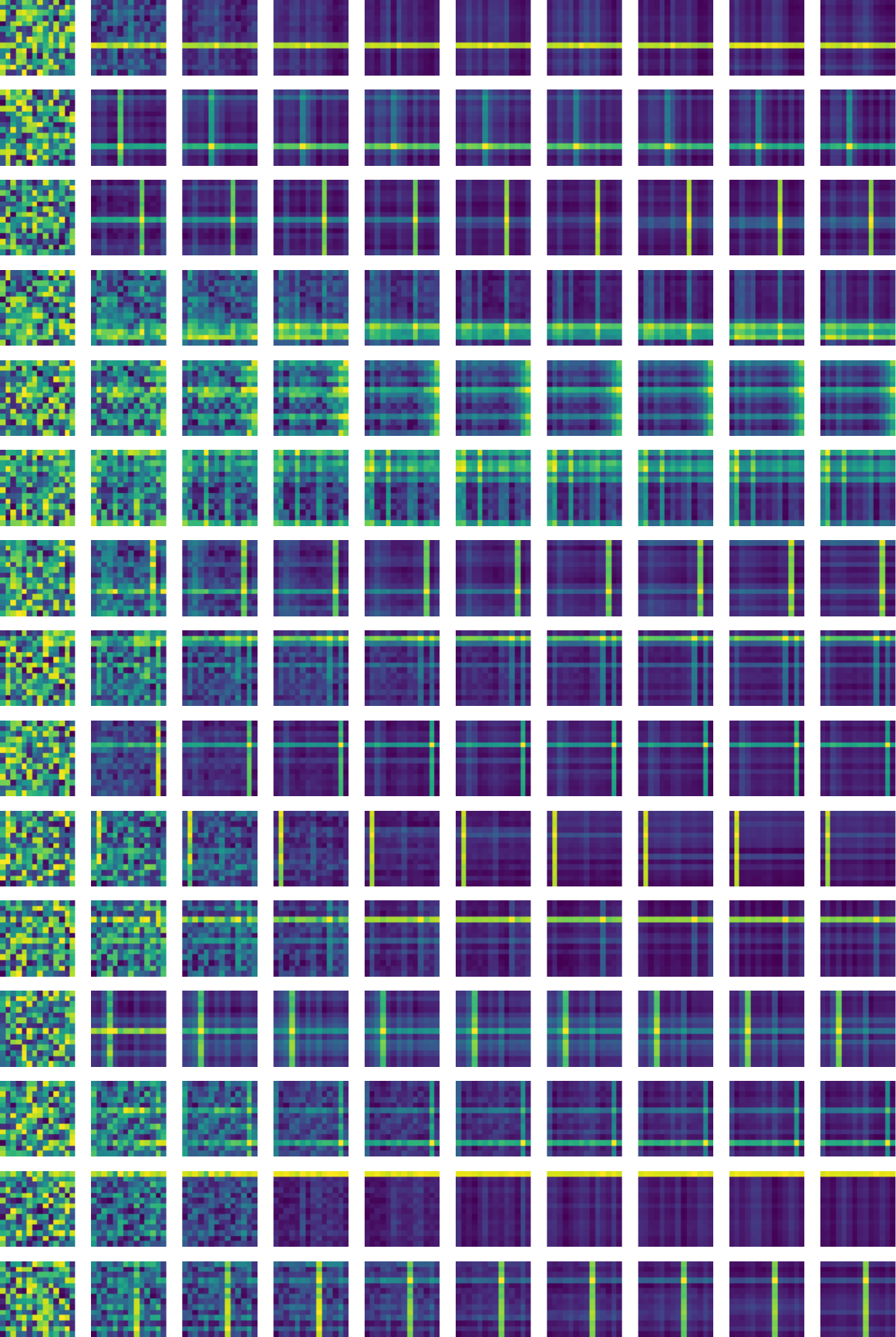}
  %\fbox{\rule[-.5cm]{0cm}{4cm} \rule[-.5cm]{4cm}{0cm}}
  \caption{Discrete model feed-forward receptive fields evolution for the crossbar simulation. Rows are different neurons and columns are different points in time (shown at every 1 million Euler steps)}
  \label{figure_disc_evolution_crossbar}
\end{figure}

\begin{figure}[!ht]
  \centering
  \includegraphics[width=\textwidth]{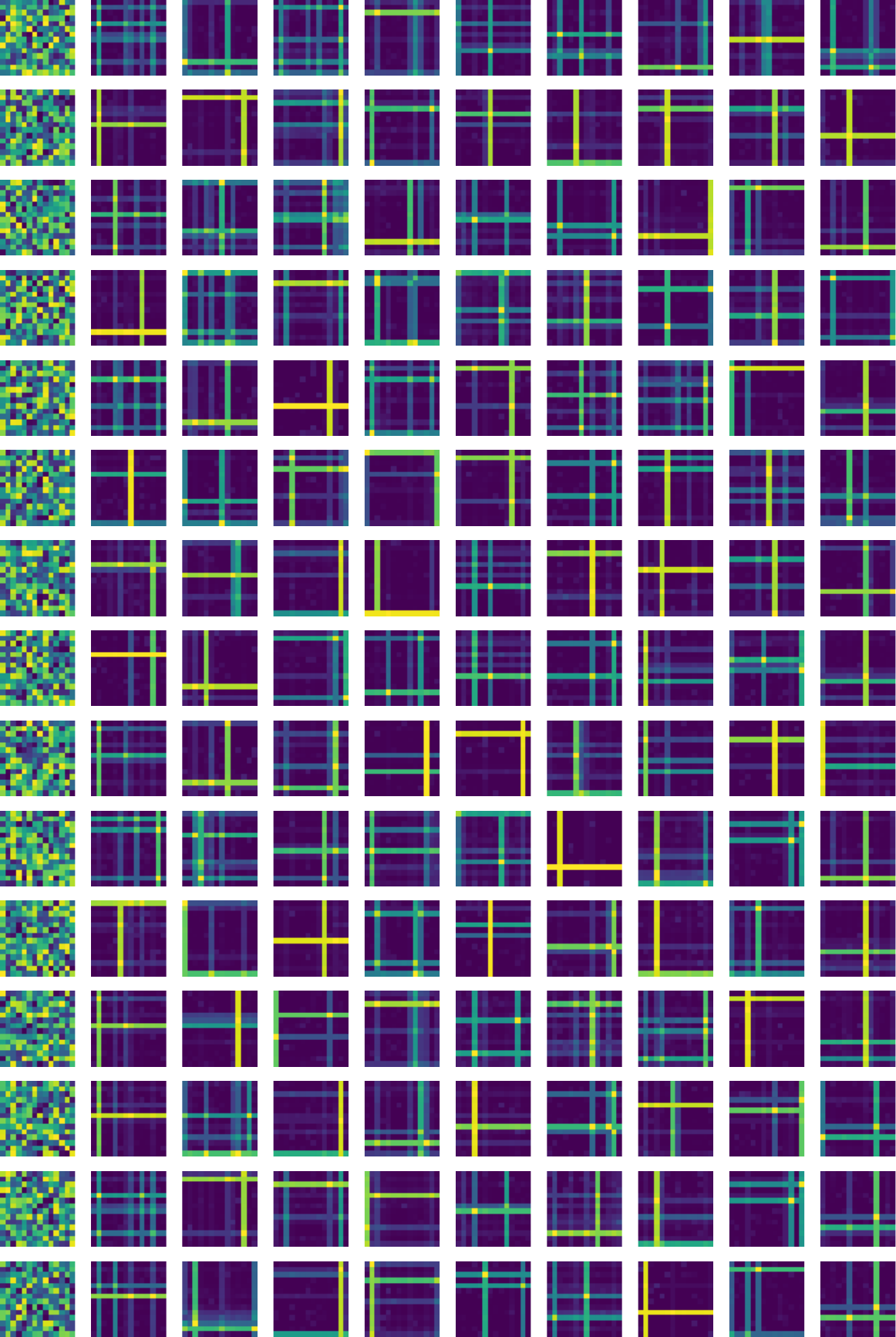}
  %\fbox{\rule[-.5cm]{0cm}{4cm} \rule[-.5cm]{4cm}{0cm}}
  \caption{Continuous model feed-forward receptive fields evolution for the crossbar simulation. Rows are different neurons and columns are different points in time (shown at every 1 million Euler steps)}
  \label{figure_conti_evolution_crossbar}
\end{figure}

\begin{figure}[!ht]
  \centering
  \includegraphics[width=\textwidth]{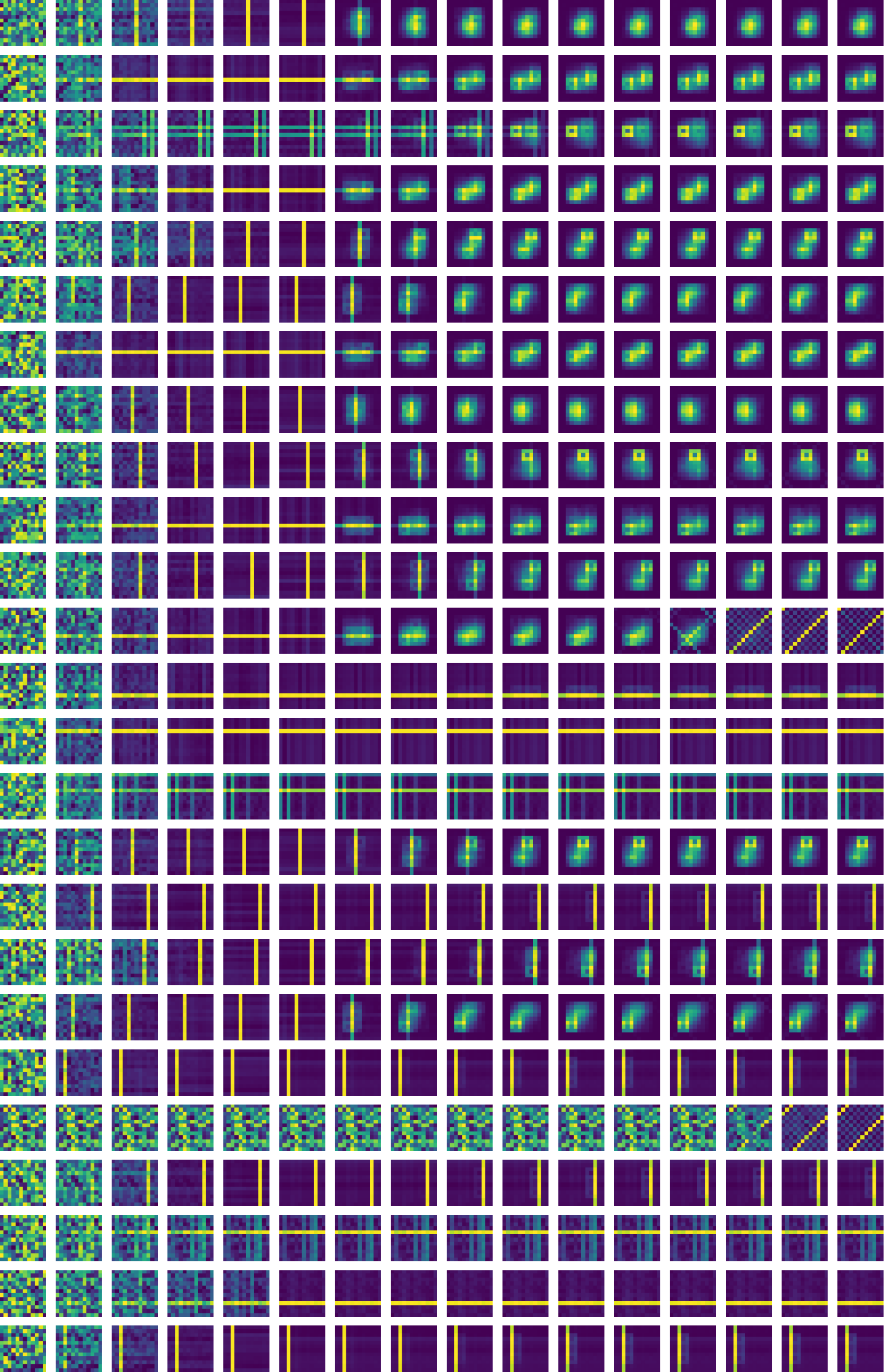}
  %\fbox{\rule[-.5cm]{0cm}{4cm} \rule[-.5cm]{4cm}{0cm}}
  \caption{Asynchronous model feed-forward receptive fields evolution for the mixed stimuli simulation. Rows are different neurons and columns are different points in time (shown at every 1 million Euler steps)}
  \label{figure_async_evolution_mixed_stimuli}
\end{figure}

\begin{figure}[!ht]
  \centering
  \includegraphics[width=\textwidth]{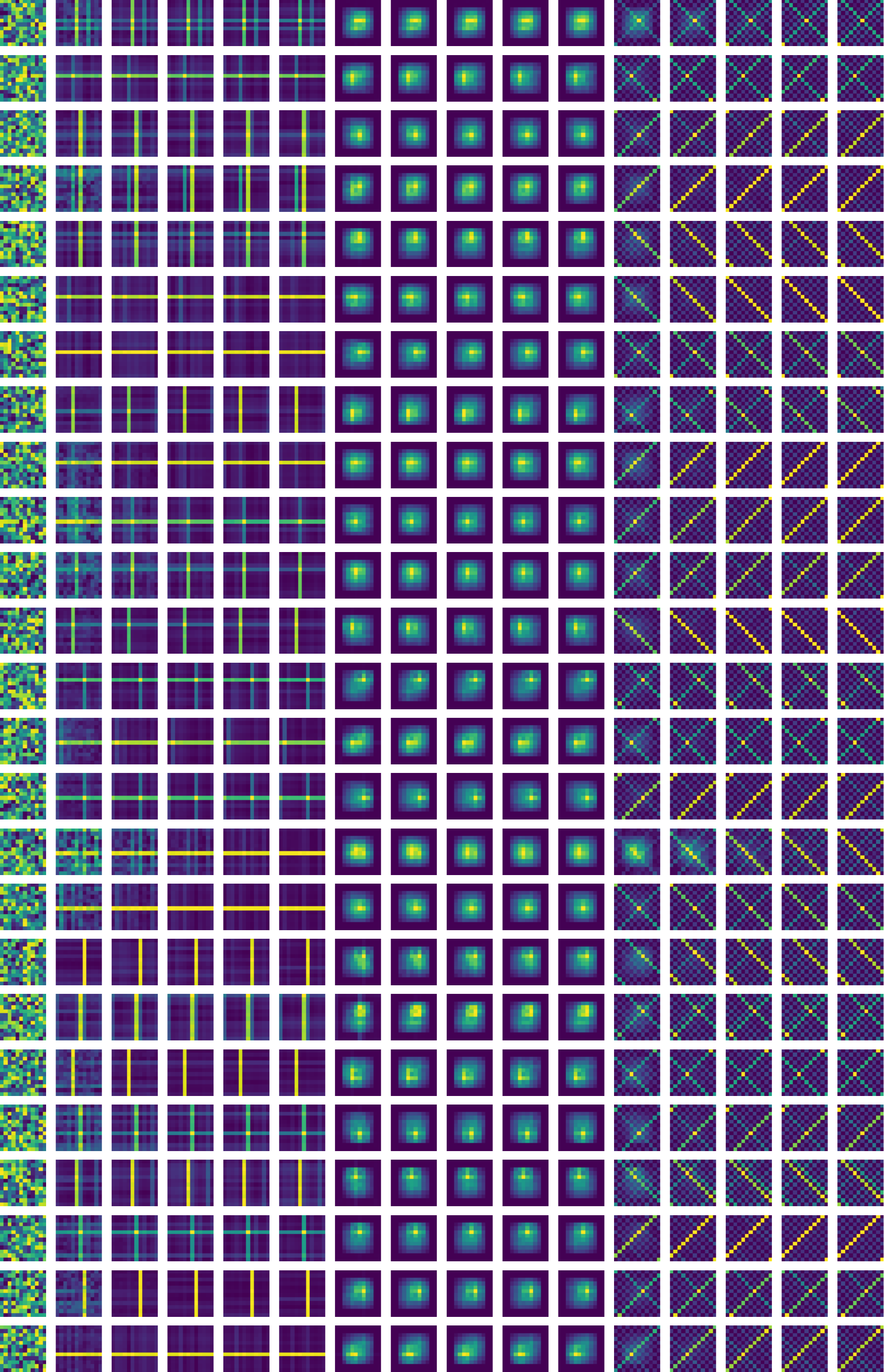}
  %\fbox{\rule[-.5cm]{0cm}{4cm} \rule[-.5cm]{4cm}{0cm}}
  \caption{Discrete model feed-forward receptive fields evolution for the mixed stimuli simulation. Rows are different neurons and columns are different points in time (shown at every 1 million Euler steps)}
  \label{figure_disc_evolution_mixed_stimuli}
\end{figure}

\end{document}